\documentclass[aps,twocolumn,prl,showpacs]{revtex4}
\usepackage{graphicx}
\usepackage{amssymb}
\usepackage{amsmath}
\usepackage{color}
\def\cpl{Chem. Phys. Lett.}

\def\pra{{Phys.~Rev.~A}}        % Physical Review A: General Physics
\def\prl{{Phys.~Rev.~Lett.}}    % Physical Review Letters
\def\jcp{{J.~Chem.~Phys.}}      % Journal of Chemical Physics
\def\jpb{{J.~Phys.~B}}      % Journal of Physics B

\begin{document}

\title{Efficient formation of  deeply bound ultracold molecules probed by broadband detection}

\author{Matthieu Viteau$^{1}$, Amodsen Chotia$^{1}$, Maria Allegrini$^{1,2}$, \\ Nadia Bouloufa$^{1}$, 
 Olivier Dulieu$^{1}$, Daniel Comparat$^1$,  Pierre Pillet$^1$}
\email{Daniel.Comparat@lac.u-psud.fr}
\affiliation{$^{1}$Laboratoire Aim\'{e} Cotton, CNRS, Univ Paris-Sud, B\^{a}t. 505, 91405 Orsay, France}
\affiliation{$^{2}$CNISM, Physics Department, Pisa University, Largo Pontecorvo, 3 56127 PISA, Italy}

\date\today

\begin{abstract}
Using a non-selective broadband detection scheme we discovered an efficient mechanism of formation of ultracold Cs$_2$ molecules in deeply bound levels ($v=1-9$) of their electronic ground state X$^1 \Sigma_g^+$. They are formed by a one-photon photoassociation of ultracold cesium atoms in a manifold of excited electronic states, followed by a two-step spontaneous emission cascade. We were able to form about $10^5-10^6$ molecules per second in these low vibrational levels of the ground state. This detection scheme could be generalized to other molecular species for the systematic investigation of  cold molecule formation mechanisms.

\end{abstract}


\maketitle

The creation and the study of ensembles of cold and ultracold molecules attracts considerable attention \cite{2004EPJD...31..149D,DulieuJPB2006,Krems2008,Krems}.
Slowing pre-existing molecules using for instance buffer-gas cooling or supersonic beam deceleration typically delivers molecules with a translational temperature down to a few millikelvins. The only way to produce molecules with a temperature in the sub-millikelvin range relies on the association of ultracold atoms. In quantum degenerate gases a magneto-association step via Feshbach resonances  \cite{2006RvMP...78.1311K,2008RvMP...80..885B} followed by an adiabatic population transfer was recently found very successful to form ultracold molecules in a single deeply bound level \cite{2008Sci...321.1062D,2008NatPh...4.....O,2008Sci_Ni}.
Besides, the photoassociation (PA) of ultracold atoms from a standard magneto-optical trap (MOT) is a well-known efficient process for the formation of ultracold molecules, with a rate as high as $\sim 10^6-10^7\,$s$^{-1}$ \cite{2006RvMP...78..483J}. The main drawback of the PA approach is the spread of the population over many vibrational levels with low binding energy. This can be partly circumvented via more elaborated  schemes
when the molecular electronic structure is well known enough. For instance, ultracold ground state K$_2$ molecules created in the lowest vibrational level $v=0$ have been observed via a two-step PA scheme \cite{nikolov2000}, while a sequence of a PA step followed by a further absorption-emission transfert  have produced ultracold ground state RbCs molecules in $v=0$ as well \cite{2005PhRvL..94t3001S}. Ultracold LiCs molecules in the absolute ground state rovibrational level $v=0, J=0$ have also been detected after a single PA step \cite{2008PhRvL_deiglmayr}. However in all these experiments the formation rate is limited to $\sim 10^3\,$s$^{-1}$ molecules in $v=0$.

Therefore a major challenge is to discover novel experimental PA schemes which could produce ultracold ground state molecules in deeply-bound levels, ultimately in $v=0, J=0$. This is a considerable task, as such schemes could generally involve peculiarities of the electronic structure of each individual molecular species: potential barrier \cite{vatasescu2000,vatasescu2006}, shape \cite{boesten1996} or Feshbach \cite{2003EL.....64..171T} resonances in the ground state, non-adiabatic couplings in the PA state \cite{2001PhRvL..86.2253D}, flux enhancement \cite{1998PhRvL..80..936G}, or accidental matching of radial wave functions related to each step of the process \cite{nikolov2000}.

In this letter, we demonstrate a general and systematic method to look for efficient ultracold molecule formation schemes based on PA, without knowing {\it a priori} the details of the molecular structure. It is based on a detection procedure which does not select the population of a particular bound level, in contrast with all previous experiments relying on a Resonantly-Enhanced Multi-Photon Ionization (REMPI) of the formed molecules \cite{2006RvMP...78..483J}. Applying this technique on a cesium MOT, we discovered a PA process resulting in a high formation rate ($\sim 10^6\,$s$^{-1}$) of deeply-bound ground state molecules, down to the $v=1$ level. The mechanism is modeled by a single-photon PA step exciting coupled molecular states, followed by a two-step spontaneous emission cascade.

\begin{figure}[htp!]
	\centering
		\resizebox{0.45\textwidth}{!}{
		\includegraphics*[87mm,50mm][165mm,165mm]{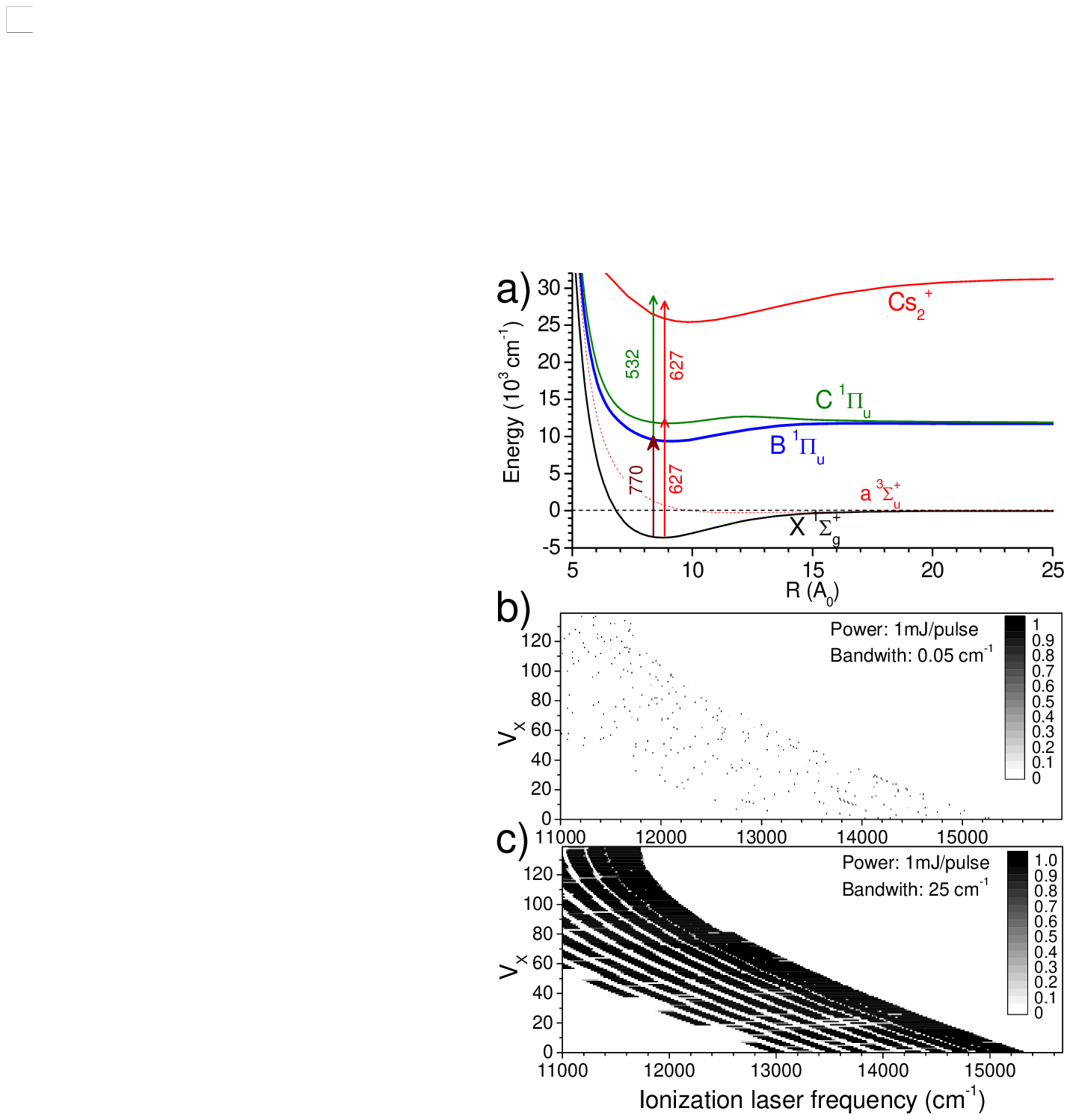}}
		\caption{(a) REMPI detection scheme of deeply-bound ground state Cs$_2$ molecules with a broad-band laser (770~nm and 532~nm) via the $B^1\Pi_u$ state, and with a narrowband laser (627~nm) via the $C^1\Pi_u$ state. (b) and (c) Comparison of transition probabilities (in level of grey) for the transitions at energy $E_{X-B}$ between  vibrational levels of the $X$ ($v_X$) and $B$ states, for a laser linewidth of $0.05\,$cm$^{-1}$ (b) and of $25\,$cm$^{-1}$ (c), with identical power (1~mJ/pulse). The probability is put to unity for a saturated transition.}
	\label{Fig_det_BB}
\end{figure}

PA of cold cesium atoms \cite{1998PhRvL..80.4402F} is achieved with a cw Titanium:Sapphire laser (intensity 300~W.cm$^{-2}$), pumped by an Argon-ion laser, exciting molecules which can decay by spontaneous emission into vibrational levels of the molecular ground state $X^{1}\Sigma_{g}^{+}$,  (hereafter referred to as X), or of the lowest triplet state $a^{3}\Sigma_{u}^{+}$. In order to observe deeply-bound molecules in the $X$ state which could result from an {\it a priori} unknown mechanism, we set up a broadband REMPI detection through vibrational levels $v_B$ of the spectroscopically known $B^1\Pi_u$ excited state \cite{1989CPL...164..419D} (refereed to as the $B$ state). The two-photon transition is induced by a pulsed dye laser (LDS751 dye, wavelength $\sim 770$~nm, pulse energy $\sim 1$~mJ, focused waist $\sim 500\,\mu$m) and by the pump laser ($532$~nm wavelength) as illustrated in Fig.\ref{Fig_det_BB}a. The formed Cs$_{2}^{+}$ ions are then detected using a pair of microchannel plates through a time-of-flight mass spectrometer.
%In our experiment \cite{1998PhRvL..80.4402F}, the cesium photoassociation step is achieved using a cw Titanium:Sapphire laser (intensity 300 W.cm$^{-2}$), pumped by an Argon-ion laser. The photoassociated molecules could decay by spontaneous emission into stable vibrational levels of the molecular ground state, $\text{a}^{3}\Sigma _{u}^{+}$ or $\text{X}^{1}\Sigma _{g}^{+}$,  hereafter referred to as X. Our present goal is to detect deeply bound singlet  X molecules formed through an efficient photoassociation process. But, without any a priori knowledge of the  process, we need to tune the detection laser to ionize most of the molecules formed in several vibrational levels.  As shown in the upper part of Fig. \ref{Fig_det_BB}), we choose a REMPI process through the deep $B^1\Pi_u$ excited state (refereed as the B state)  \cite{1989CPL...164..419D} to detect the molecules which are stabilized in vibrational levels of the X state. This REMPI  scheme  uses a dye laser  photon ( LDS751 dye, wavelength $\sim 770$ nm,  pulse energy $\sim 1\,$mJ, focused waist $\sim 500\,\mu$m) plus a photon from the  pump laser ($532 \,$nm wavelength) to ionize  the molecules. The formed Cs$_{2}^{+}$ ions are then detected using a pair of microchannel plates through a time-of-flight mass spectrometer.

%To potentially ionize  most of the formed molecules in the singlet X state,
The major advance of the present experiment compared to previous ones is the broadband detection of the formed ultracold molecules. We replaced the grating in the pulsed dye laser cavity by a less dispersive prism, which broadens its linewidth from $\sim 0.05$~cm$^{-1}$ to $\sim 25$~cm$^{-1}$. We display in Fig.\ref{Fig_det_BB} the results of a modeling of the ionization process, for both narrowband (panel b) and broadband (panel c) schemes. We assume that the ionization probability due to second ($532$~nm) photon is independent of the specific $v_B$ level, so that it is proportional to the population of this level induced by the first photon at $770$~nm. The excitation probabilities of the $v_X$ levels toward the $v_B$ levels are obtained from Franck-Condon factors computed for the experimentally known $X$ and $B$ potential curves \cite{1985JChPh..82.5354W,1989CPL...164..419D}, assuming a constant dipole transition moment. As expected, the narrowband ionization scheme allows
  the ionization of a single $v_X$ level at a given frequency (Fig.\ref{Fig_det_BB}b). In contrast, the broadband scheme involves a laser pulse width of the order of the vibrational spacing of both the $X$ and $B$ states (up to 40~cm$^{-1}$), so that many vibrational $v_X$ levels can be ionized in a single shot (Fig.\ref{Fig_det_BB}c). For instance, a laser pulse at $\sim 11730$~cm$^{-1}$ or at $\sim 13000$~cm$^{-1}$ would excite almost all molecules lying in vibrational levels  $v_X>37$ or $v_X<70$ respectively.

\begin{figure}[htp!]
	\centering
\resizebox{0.48\textwidth}{!}{
		\includegraphics*[15mm,116mm][180mm,218mm]{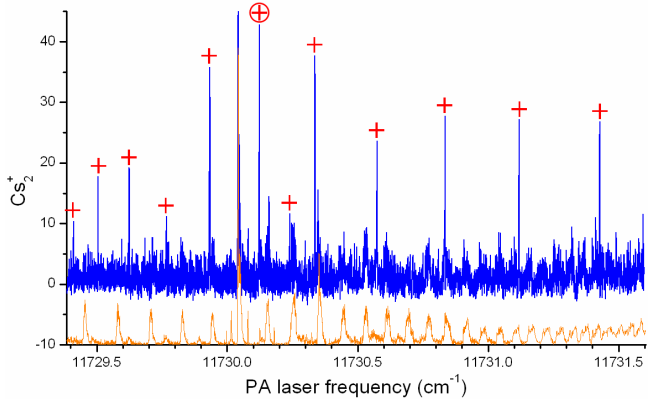}}
				\caption{Upper trace: Cs$_2^+$ ion spectrum recorded after scanning the frequency of the PA laser below the $6s+6p_{3/2}$ dissociation limit, and using the broadband REMPI detection laser with energy around 13000~cm$^{-1}$. The crosses label the previously unobserved PA lines. Lower trace: Cs$_2^+$ ion spectrum obtained using the conventional narrowband REMPI detection \cite{1998PhRvL..80.4402F}, displayed with an offset of 10 ions for clarity.}
	\label{FigSBB}
\end{figure}

Choosing the latter energy range for the broadband detection laser ($\sim 13000$~cm$^{-1}$) we scanned the PA laser frequency over a few wavenumbers below the $6s+6p_{3/2}$ dissociation limit. We discovered several intense PA lines labeled with crosses in Fig.\ref{FigSBB}, revealing a large number of ultracold molecules formed in low ($v_X < 70$) vibrational levels of the $X$ state. These detected singlet molecules were present in our previous experiments performed in the same PA energy range \cite{1998PhRvL..80.4402F}, but our previous narrowband REMPI detection scheme (wavelength $\sim 720\,$nm)
was optimized to detect triplet $\text{a}^{3}\Sigma_{u}^{+}$ molecules, and therefore was blind for these singlet molecules (see lower spectrum of Fig. \ref{FigSBB}).

\begin{figure}[htp!]
	\centering
		\resizebox{0.48\textwidth}{!}{
		\includegraphics*[37mm,170mm][163mm,270mm]{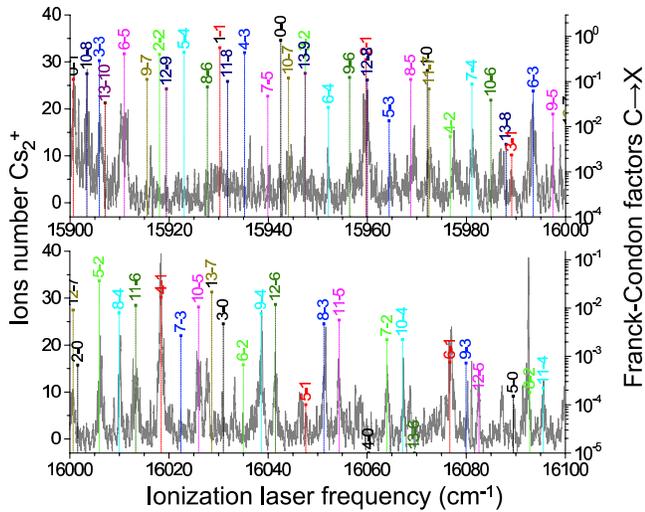}}
		\caption{Cs$_2^+$ ion count (left vertical axis) resulting from a narrowband REMPI detection (frequency $\sim 627$~nm). The PA laser energy is fixed at $11730.1245$~cm$^{-1}$ corresponding to the position of the most intense line marked with a circled cross in Fig. \ref{FigSBB}.  Transition labels $v_C-v_X$ are extracted from computed Franck-Condon Factors (right vertical axis) between vibrational levels of the spectroscopically known $C$ and $X$ states.}
	\label{FigSXC}
\end{figure}

To determine precisely the internal state of these formed molecules we performed conventional REMPI spectra using a narrowband (DCM) dye laser (frequency $\sim 627$~nm, see Fig.\ref{Fig_det_BB}a). We fixed the PA laser energy on the most intense line of Fig.\ref{FigSBB} (at $11730.1245$~cm$^{-1}$), about $2$~cm$^{-1}$ below the $6s+6p_{3/2}$ asymptote. Then we scanned the ionization laser frequency to record the ionization spectrum of the ground state molecules excited through the intermediate $\text{C}^1\Pi_u$ state (Fig.\ref{FigSXC}). Taking advantage of the spectroscopic knowledge of the $X-C$ transitions \cite{1985JChPh..82.5354W,1982JChPh..76.4370R}, the lines were easily assigned to transitions from ground state vibrational levels restricted to the range $v_X=1$ to $v_X=9$.  Taking into account the efficiency of the detection \cite{1998PhRvL..80.4402F}, the ion signal corresponds to a cumulated formation rate for the $v_{\text{X}}<10$ molecules close to 10$^{6}$ per second.

\begin{figure}[htp!]
	\centering
	\resizebox{0.48\textwidth}{!}{
		\includegraphics*[6mm,6mm][118mm,63mm]{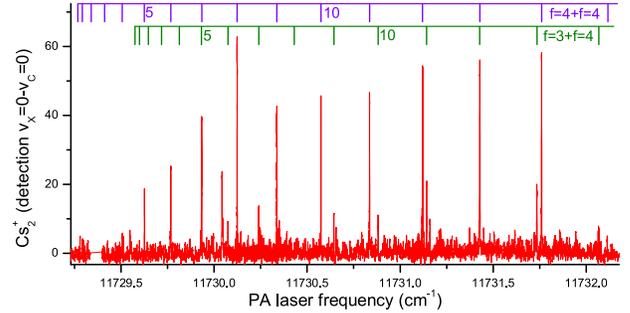}}		
        \caption{Same as Fig. \ref{FigSBB} with an additional vibrational cooling step  \cite{MatthieuViteau07112008}. Cs$_2^+$ ions are detected with a narrowband laser (wavenumber 15941~cm$^{-1}$) tuned to resonance with the transition $(v_X=0) \rightarrow (v_C=0)$. An intense series and a weak rotational series, separated by 9.2~GHz have been fitted, assigning the lines to $J=5$ to 14. Extrapolated line positions for $J=0$ to 4 are also displayed.}
	\label{FigSRot}
\end{figure}

\begin{figure}[htp!]
	\centering
	%\rotatebox{-90}{
	\resizebox{0.48\textwidth}{!}{
		\includegraphics*[46mm,57mm][179mm,149mm]{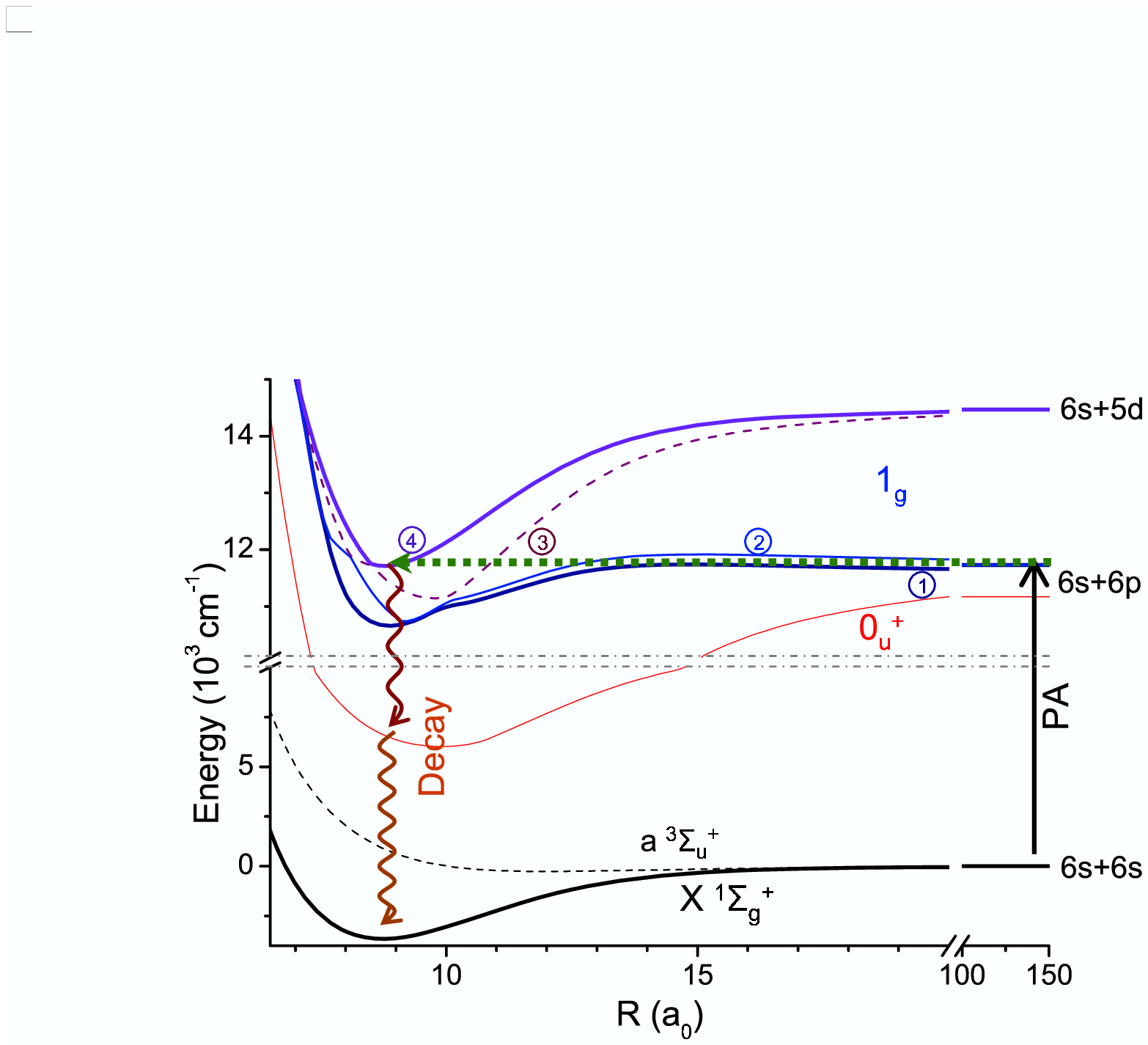}}%}
				\caption{Theoretical Cs$_2$ molecular potential curves including spin-orbit \cite{deiglmayr_unpublished,lyon_private}, relevant for the present PA and cold molecule formation process. The PA laser excites levels of a long-range 1$_{g}$ curve (label 1), which is coupled to the $v=0$ level of the short range 1$_{g}$ curve (label 4) through several avoided crossings involving $1_g$ curves labeled 2 and 3. Formation of deeply-bound ground state molecule proceed through  a spontaneous emission cascade via the $0_{u}^{+}$ states.}
	\label{Meca}
\end{figure}

In order to further investigate this novel efficient PA mechanism, we improved the PA signal of Fig. \ref{FigSBB} by (i) performing a vibrational cooling step which accumulates the $v_{\text{X}}<10$ population into the sole $v_X=0$ level \cite{MatthieuViteau07112008}, and (ii) by detecting these molecules using the narrowband detection through the known transition between X$(v_X=0)$ and C$(v_C=0)$. The PA spectroscopy performed under such conditions is depicted in Fig.\ref{FigSRot}. Two series of lines separated by $9.2$~GHz (i.e. the hyperfine structure of Cs$(6s_{1/2})$) are visible. As the MOT mainly contains Cs($f=4$) atoms, the intense lines correspond to PA of two Cs$(f=4)$ atoms, and the weak lines to PA of one Cs$(f=4)$ and one Cs$(f=3)$ atom. This $9.2$~GHz line spacing rules out the possibility that the molecules are formed after a  PA step with two identical photons, as this would induce a twice small spacing. Furthermore the series of intense lines is easily assigned to a rotational progression with rotational quantum number between $J=5$ to $14$. We fitted a rotational constant $B_v=0.01188(1)$~cm$^{-1}$, corresponding to an approximate average internuclear distance $\bar{R}_e \equiv (2 \mu B_v)^{-1/2}=8.73a_0$, where $\mu$ is the Cs$_2$ reduced mass ($a_0=0.0529177$~nm). Thus this novel single-photon PA mechanism excites both (Cs$(f=4)$, Cs$(f=4)$) and (Cs$(f=4)$, Cs$(f=3)$) ground state atom pairs, in a level located about 2~cm$^{-1}$ below the $6s+6p_{3/2}$ asymptote, with a vibrational motion taking place mainly in the short distance range. Relying on theoretical Cs$_2$ potential curves including spin-orbit \cite{deiglmayr_unpublished,lyon_private}, we identified only one excited potential curve, belonging to the so-called $1_g$ symmetry, as a good candidate for the PA state. We depict the process as follows: the PA laser excites the atom pair into a bound level of the lowest $1_g(6s+6p_{3/2})$ long-range potential curve (curve 1 in Fig.\ref{Meca}), which is coupled at short distances to the lowest $1_g(6s+6d_{5/2})$ potential curve (curve 4 in Fig.\ref{Meca}), through several avoided crossings induced by spin-orbit interaction. The $v=0$ level of the curve 4 is predicted with an energy very close to the $6s+6p_{3/2}$ dissociation energy. It is most probably the only populated short-range level in this mechanism, which then decays down to the $X$ ground state through a two-photon spontaneous emission cascade via the $0_{u}^{+}$ potentials. Note that the spontaneous decay cannot proceed down to the levels of the $a^{3}\Sigma_{u}^{+}$ state, as the average distance $8.73a_0$ corresponds to the range of the repulsive wall so that only dissociating pairs could be formed. Finally, it is worthwhile to mention that a similar situation involving internal couplings has already been demonstrated for the $0_u^+$ molecular symmetry in Cs$_2$ \cite{2001PhRvL..86.2253D}.

%With these points we can then assume that we photoassociate directly with only one photon (starting from  Cs$(6s_{1/2})f=3+$Cs$(6s_{1/2})f=3$ or  Cs$(6s_{1/2})f=3+$Cs$(6s_{1/2})f=4$ collision) just below the $6s-6p_{3/2}$ asymptote and in a potential with a centroid  at $R_e=8.15a_0$. Only one potential respects this parameters: the $1_g (6s+5d)$ potential. All $1_g (6s+6p)$ and $1_g (6s+5d)$ potentials are presented in Fig.\ref{Meca} indicating that the long-range radial wavefunction is coupled to short range radial wavefunction by internal (spin-orbit) coupling of the potentials. Similar coupling, between long and short-range wavefunctions, have already been observed in $0_u^+$ potentials \cite{2001PhRvL..86.2253D}. Therefore, the photoassociation laser couples the colliding atoms wavefunction to the long-range radial wavefunction; the spontaneous decay is then possible due to an efficient coupling at internuclear distance near $8.15\,a_0$. In order to preserve the gerade $g$, ungerade $u$ symmetry, we know (see  Fig. \ref{Meca}) that the ground-state molecules, X$^{1}\Sigma _{g}^{+}(v=1-9)$, are therefore formed, from the $1_g$ state, by a two photon spontaneous emission cascade via the $0_{u}^{+}$ potentials. At $\sim 8.15a_0$ internuclear distance, the single photon decay toward the a$^{3}\Sigma _{u}^{+}$ potential occurs in the continuum and do not lead to molecule formation.

To investigate the reason for the high values of the rotational quantum number, up to $J=14$, of the PA molecules we turned off the MOT lasers 2~ms before switching on the PA laser, and while the PA laser was on (1 ms). We observed no change in the PA spectrum. This demonstrates that the MOT lasers (both trapping and repumping ones) do not bring additional angular momentum into the process, in contrast to previous observations in sodium PA (up to $J=22$) \cite{2001PhRvA..63b1401S}, or in cesium PA (up to $J=8$) \cite{fioretti1999a}. Such high $J$ values are probably induced by the large hyperfine structure of the long-range $1_g(6s+6p_{3/2})$ state \cite{2006RvMP...78..483J}. A strong mixing between hyperfine and rotational structure is expected, just like in the well studied $1_u(6s+6p_{3/2})$ Cs$_2$ molecular state \cite{2000EPJD...11...59C}. 
The $1_g(6s+6p_{3/2})$ levels are characterized by a value of the total angular momentum ${\bf F= J + I}$, where $\bf I$ is the total nuclear angular momentum  ($I\leq 7$) of a cesium atom pair. The PA excitation of non-rotating ground state atom pairs, which takes place at large distances, creates a strongly mixed hyperfine-rotational level of the $1_g(6s+6p_{3/2})$ state with  $F \sim I \leq 7$. Due to the high density of levels near the dissociation threshold, this PA level  can be efficiently coupled to several rotational states of the $v=0$ level of the internal $1_g(6s+6d_{5/2})$ potential curve. The rotational angular momentum ${\bf J=F - I}$ can reach eigenvalues up to $J=14$ or more. The abrupt cut-off of the observed rotational series at $J=14$ in Fig.\ref{FigSRot} occurs as the next rotational levels lie above the $6s+6p_{3/2}$ dissociation limit.

In this paper we demonstrated the ability of a new broadband ionization procedure to detect most of the bound ground state molecules formed in a cold gas. This approach, combined with PA in excited electronic states, provides a general method for the search of novel paths for formation  of cold molecules.
Applied to a Cs MOT, this novel approach allowed us to detect deeply bound Cs$_2$ molecules in the X$^1 \Sigma_g^+$ state. Combined with our recently demonstrated optical pumping technique using a tailored broadband light source, we were able to form about $10^5-10^6$ molecules per second in the $v=0$ level of the ground state \cite{MatthieuViteau07112008}. The  simplicity of the experiment (one-step photoassociation) contrasts with the complexity of the interpretation of the  photoassociation process which involves, rotational, hyperfine and spin-orbit coupling of four potential curves. The formation process also revealed an unexpected  two-photon spontaneous emission cascade responsible for the molecule formation. This scheme opens the possibility, to stimulate the first photon of the cascade to enhance the cold molecules formation rate. In the future such molecules could be accumulated in an optical trap to study collisional processes between cold atoms and molecules in or
 der to assess the efficiency of evaporative cooling, or to investigate ways for achieving controlled chemistry or observing dipolar interactions in ultracold gases.

This work is supported by the "Institut Francilien de Recherche sur les Atomes Froids" (IFRAF). M.A. thanks the EC-Network EMALI.

%\bibliography{Bib_Mol_broad}

\end{document}